\documentclass[prl,twocolumn,superscriptaddress]{revtex4-1}
\pdfoutput=1
\usepackage[letterpaper,top=1.45cm,bottom=1.65cm,left=2cm,right=2cm]{geometry}
\usepackage{amssymb,amsmath,amsthm,graphicx,mathptmx}
\usepackage{subfigure}
\usepackage{graphics, color}
\usepackage{latexsym}
\usepackage{bm}
\usepackage{epsfig}
\usepackage{multirow,tabularx}
\usepackage[colorlinks=true,linktocpage=true,linkcolor=blue,citecolor=blue]{hyperref}
\usepackage[abs]{overpic}

\usepackage{fancyhdr}
\fancypagestyle{plain}

\newcommand{\PRLsection}[1]{\emph{#1.---}}

\newcolumntype{Y}{>{\centering\arraybackslash}X}

\begin{document}

\title{Bare Quantum Null Energy Condition}

\author{Zicao Fu}
\email{zicaofu@physics.ucsb.edu}
\affiliation{Department of Physics, University of California, Santa Barbara, California 93106, USA}

\author{Donald Marolf}
\email{marolf@physics.ucsb.edu}
\affiliation{Department of Physics, University of California, Santa Barbara, California 93106, USA}

\begin{abstract}
The quantum null energy condition (QNEC) is a conjectured relation between a null version of quantum field theory energy and derivatives of quantum field theory von Neumann entropy. In some cases, divergences cancel between these two terms and the QNEC is intrinsically finite. We study the more general case here where they do not and argue that a QNEC can still hold for bare (unrenormalized) quantities. While the original QNEC applied only to locally stationary null congruences in backgrounds that solve semiclassical theories of quantum gravity, at least in the formal perturbation theory at a small Planck length, the quantum focusing conjecture can be viewed as the special case of our bare QNEC for which the metric is on shell.
\end{abstract}

\maketitle

\PRLsection{Introduction}
Motivated by features desired when
relativistic quantum field theories (QFTs) are coupled to
gravity, in Ref. \cite{Bousso:2015mna}, Bousso, Fisher, Liechenauer, and Wall
proposed the quantum null energy condition (QNEC),
which states that the (null) energy density at a point is
bounded below by a second derivative at that point of an
appropriate von Neumann entropy. Schematically, the
relation takes the form
\begin{equation}
\label{QNEC}
Q:=T_{ab}k^ak^b-\frac{1}{2\pi }S''\ge 0,
\end{equation}
where $k^a$ is the generator of the congruence $N$, $T_{ab}$ is the stress tensor, and $S$ is the von Neumann entropy on one side of the null congruence $N$. The original conjecture assumed the null congruence to be locally stationary at the desired point $p$, meaning that the expansion $\theta$ and shear $\sigma_{\alpha \beta}$ satisfy $\sigma_{\alpha \beta}|_p =0$ and $\theta|_p = \dot{\theta}|_p = 0$, where the overdot indicates a derivative with respect to an affine parameter along the generator through $p$ and greek indices $\left( \alpha ,\beta ,\gamma ,\ldots \right)$ run over the $d-2$ directions associated with the cut of the null congruence at which $S''$ was computed. With this restriction \cite{footnote1}, \eqref{QNEC} has been shown to hold in Minkowski space for both free theories \cite{Bousso:2015wca} and interacting QFTs that flow to a nontrivial conformal fixed point in the ultraviolet \cite{Balakrishnan:2017bjg}. However, we will drop the locally stationary condition below and will also consider general curved spacetimes.

Now, both terms in \eqref{QNEC} diverge for continuum QFTs. But the divergences cancel in flat spacetime, on Killing horizons, and also more generally when certain dimension-dependent derivatives of $\theta$ and $\sigma _{\alpha \beta }$ vanish \cite{Fu:2017evt,Akers:2017ttv}. In such cases, the quantity $Q$ in \eqref{QNEC} is intrinsically finite and may be equivalently expressed in terms of either any renormalized $T_{ab}$ and $S$ or in terms of the corresponding bare quantities. The only restriction is that one use the same regulator and/or renormalization scheme to define both $T_{ab}$ and $S$, say by computing both from a common regulated or renormalized partition function, varying the background metric to obtain $T_{ab}$ and using the replica trick to define $S$. With the above understanding, the difference $Q$ is then also independent of the choice of regulator or renormalization scheme. We thus describe such contexts by saying that the QNEC \eqref{QNEC} is scheme-independent. In contexts where this property is known to hold, the QNEC was proven in Refs. \cite{Fu:2017evt,Akers:2017ttv} for the universal sector of holographic QFTs in $d=2,3,4,5$ spacetime dimensions at leading order in small bulk Newton constant $G_N$ (and, when appropriate, also small string length $\ell_{\rm s}$).

Restricting the QNEC to scheme-independent contexts is natural from the point of view of Ref. \cite{Bousso:2015mna},  as
this work motivated the QNEC by studying the weak
gravity limit of another conjecture, the quantum focusing conjecture (QFC), which involved intrinsically
finite quantities. Since the QFC is always scheme
independent, it can have only scheme-independent
consequences. The extra assumptions needed to derive
the QNEC from the QFC must thus guarantee scheme
independence for the QNEC. However, at least in
holographic theories, it was later noted in Ref. \cite{Fu:2017evt}  that
the QNEC holds in contexts where it cannot be derived
from the QFC. Taking this as inspiration, we consider
arbitrary curved backgrounds below as well as general
choices of N. We thus explicitly allow the possibility
that divergences in \eqref{QNEC} fail to cancel.

In scheme-dependent contexts, the physical content of
conjecturing \eqref{QNEC} then depends on whether one uses bare
or renormalized quantities and also on the choice of
regulator and of any renormalization scheme. For inspiration, we thus turn to more familiar inequalities such as
positivity of entropy or strong subadditivity. When such
inequalities become scheme dependent, they typically
hold only for the bare quantities and even then only
when defined by an appropriately physical cutoff such as
putting the theory on a lattice. Removing the regulator
yields a divergence but of the physically correct sign. For
example, in quantum mechanics, entropy itself is an
intrinsically positive quantity. And using a lattice regulator to define the von Neumann entropy of a region in
some continuum field theory gives a positively signed
infinite result, but see, e.g., \cite{Donnelly:2014fua} for a discussion of other
regulators that can yield negative entropies. While such
positive divergences may make the inequalities appear
trivial in the continuum limit, they can still be useful.

Following this model, we take the inequality \eqref{QNEC} to be defined by bare quantities associated with an appropriately physical regulator below. We then show for $d\le 4$ that the divergent parts of \eqref{QNEC} are always of the right sign. The details of the regulator turn out to be irrelevant, and in fact our analysis depends only on the assumption that it gives divergent-but-positive von Neumann entropies (specifically, with a UV divergence following an area law). Thus the bare QNEC holds for such regulators if the arguments of Ref. \cite{Balakrishnan:2017bjg} can be extended from flat space to the general scheme-independent context. We also show for $d\le 5$ that a suitably smeared version holds for leading-order holographic theories with Fefferman-Graham regulators. In all cases, we assume that the dynamics at the point $p$ used to evaluate \eqref{QNEC} depends on the spacetime location only through the background metric $g_{ab}$ or, equivalently but in language that may be more familiar to some readers, that all nonmetric background fields are constant scalars. More general background fields at $p$ will be saved for future investigation. However, one is free to turn on arbitrary sources at points separated from $p$, and, in particular, one may use such sources to prepare an arbitrary state of the field theory \cite{newfootnote1} at $p$.

\PRLsection{The bare QNEC}
As is well known, for any local QFT the divergences in the associated partition function ${Z}$ are integrals of local quantities. To classify such divergences, we assume our QFT to flow to some conformal fixed point in the UV and take the divergences to be built only from scalar operators and the metric. The absence of other background fields is implied by the conditions at the end of the introduction, so this will be the case when nonscalar operators have sufficiently large conformal dimensions $\Delta$. For $d\le 5$, divergent terms in the effective action must have dimension 5 or less and so are limited to the following terms:
\begin{equation}
\label{Idiv}
\begin{aligned}
I_{\rm div} = &\frac{1}{16\pi G_N} \int d^dx\sqrt{|g|}
\Bigl[\lambda_{\rm CC}{\cal O}_{\rm CC}+ \lambda_{\rm EH} {\cal O}_{\rm EH}{R} \\ & + \lambda_{R^2} R^2 + \lambda_{\rm Ricci^2} R_{ab}R^{ab} \\ & + \lambda_{\rm GB} \left( R^2 -4 R_{ab}R^{ab}+R^{abcd} R_{abcd} \right) \Bigr],
\end{aligned}
\end{equation}
where $\sqrt{|g|}$ is the metric volume element, each $\lambda$ is a (possibly divergent) constant, and ${\cal O}_{\rm CC}$ and ${\cal O}_{\rm EH}$ are local scalar operators \cite{newfootnote2}. Our convention is to express each $\lambda $ and ${\cal O}$ in terms of the large energy scale $\Lambda$ defined by the regulator. Furthermore, $R$, $R_{ab}$, and $R_{abcd}$ are the Ricci scalar, Ricci tensor, and Riemann tensor, respectively, of the spacetime metric $g_{ab}$ using the conventions of Ref. \cite{Wald:1984rg}. Up to $\log $ terms and subleading terms, we have
\begin{equation}
\label{lambdas}
\lambda_{\rm CC} \sim \Lambda^d\text{, }\lambda_{\rm EH} \sim \Lambda^{d-2}\text{, and }\lambda_{R^2}, \lambda_{\rm Ricci^2}, \lambda_{\rm GB} \sim \Lambda^{d-4}.
\end{equation}

Because unitarity bounds restrict the dimension of nontrivial scalar operators to be larger than $[(d-2)/2]$, for $d\le 5$ the 4-derivative terms must be $c$ numbers. For $d=2$, we take the UV fixed point to contain no free fields in order to forbid operators of dimension zero. Since all nontrivial operators have positive dimension, the leading term in each ${\cal O}$ is a $c$ number. It is thus a scalar background field. We have assumed such fields to be constant, so we may take each ${\cal O}$ to be of the form $a + \tilde O$, where $a_{\rm CC}$ and $a_{\rm EH}$ are finite ($c$-number) constants, $\tilde O \rightarrow 0$ as $\Lambda \rightarrow \infty$, and $\tilde{ O}$ is a sum of operators having a positive conformal dimension at the UV fixed point.  In other words, each ${\cal O}$ is the sum of a state-independent leading term and state-dependent corrections that suppressed in the limit $\Lambda \rightarrow \infty$.

The divergent contributions to \eqref{QNEC} must come from terms in \eqref{Idiv}. We analyze each below and show them to satisfy \eqref{QNEC} in at least $d\le 4$ spacetime dimensions for regulators giving $S$ the positive area-law divergence required above. Together with known results in scheme-independent contexts, this suffices to establish the claims made in the introduction.

\PRLsection{$d=2,3$}For $d=2,3$, only the cosmological constant (CC) and Einstein-Hilbert (EH) terms in \eqref{Idiv} can diverge. As described in detail in Ref. \cite{Fu:2017evt}, the cosmological constant term does not contribute to either $T_{kk}$ or $S$. In contrast, the Einstein-Hilbert ($\lambda_{\rm EH}$) term contributes to both $S$ and $T_{kk}.$ As the leading divergence which contributes to $S$, the Einstein-Hilbert term plays a special role. The contribution to $S$ for a region $A$ is proportional to the integral of $\lambda_{\rm EH}{\cal O}_{\rm EH}$ over the boundary of $A$, so, since our regulator must give a positive area-law divergence, $S>0$, we must have $\lambda_{\rm EH}{\cal O}_{\rm EH} \rightarrow a_{\rm EH} > 0$.

From Ref. \cite{Fu:2017evt}, the Einstein-Hilbert contribution to $Q$ is \cite{footnote2}
\begin{equation}
\label{EH}
\Delta Q_{\rm EH} = 2 \lambda_{\rm EH}{\cal O}_{\rm EH} \left(\frac{\theta^2}{d-2} + \sigma_{\alpha \beta} \sigma ^{\alpha \beta}\right)-2\lambda_{\rm EH}\dot{\cal O}_{\rm EH}\theta ,
\end{equation}
where contractions on $\alpha,\text{ } \beta, \text{ and } \gamma$ include projections onto the relevant $(d-2)$-dimensional surface. Since $\dot{\cal O}_{\rm EH} = \dot{\tilde{O}}_{\rm EH} \rightarrow 0$ as $\Lambda \rightarrow \infty$, the contribution $\Delta Q_{\rm EH}$ is positive at large $\Lambda$ whenever $\theta$ or $\sigma_{\alpha \beta}$ fail to vanish. And since the final term vanishes for $\theta =0$, it is non-negative in all cases. Thus any divergent parts of $Q$ satisfy \eqref{QNEC} and it remains only to check settings where $\theta $ and $\sigma_{\alpha \beta}$ both vanish so that the QNEC is scheme independent. From Refs. \cite{Bousso:2015wca,Balakrishnan:2017bjg}, we know \eqref{QNEC} to be satisfied in such cases in flat space, and Ref. \cite{Fu:2017evt} establishes \eqref{QNEC} for leading-order holographic theories with vanishing $\theta $ and $\sigma_{\alpha \beta}$ in any background. Given the similarity of Ref. \cite{Balakrishnan:2017bjg} to the flat-space holographic argument of Ref. \cite{Koeller:2015qmn}, we expect that the techniques of Ref. \cite{Balakrishnan:2017bjg} will allow this to be established for general QFTs having nontrivial UV fixed points. It would be interesting to attempt to extend the free-field light-sheet analysis of Ref. \cite{Bousso:2015wca} in a similar way.

\PRLsection{$d=4$}The analysis of the cosmological and Einstein-Hilbert terms remains as above in all higher dimensions. But the $R^2$ and Ricci-squared terms become relevant for $d=4$. While the Gauss-Bonnet term in \eqref{Idiv} can also now diverge, for $d=4$ it is a topological invariant. It thus cannot contribute to $T_{kk}$. The contribution to $S$ turns out to be topological as well, so there is no contribution to $S''$.

The $R^2$ ($\lambda_{R^2}$) term contributes
\begin{equation}
\label{R2}
\Delta Q_{R^2} = 4 \lambda_{R^2}R\left( \frac{\theta^2}{d-2} + \sigma_{\alpha \beta} \sigma ^{\alpha \beta} \right) -4 \lambda_{R^2} \dot{R}\theta ,
\end{equation}
where \eqref{R2} corrects Ref. \cite{Fu:2017evt} in the case where their $\phi_2$ depends on $R$. Note that we can ignore \eqref{R2} as when either $\theta $ or $\sigma_{\alpha \beta}$ fail to vanish, from \eqref{lambdas} it is then negligible when compared with the leading $\Lambda$ behavior of \eqref{EH}.

For the same reason, we need only compute the Ricci-squared contribution when $\theta $ and $\sigma_{\alpha \beta}$ both vanish. Now, Ref. \cite{Akers:2017ttv} computed the contribution of this term to $S''$ for $\theta =0$ and $\sigma_{\alpha \beta}=0$ and expressed the result in terms of $T_{kk}$. Moving this term to the left-hand side of the equation gives the contribution to $-Q$. And though Ref. \cite{Akers:2017ttv} used a so-called equation of motion, this was simply \cite{footnote3} $T_{kk} = k^a k^b [-2/(\sqrt{|g|})] (\delta S)/(\delta g^{ab})$, which for us is just the definition of $T_{kk}$. We can thus read off the desired result from their Eq. (3.11) to find
\begin{equation}
\label{eq:4dRicciSquared}
\begin{aligned}
\Delta Q_{\rm Ricci^2}= & 2\lambda _{\rm Ricci^2}\left[ \left(D_\alpha \theta + R_{\alpha k}\right)\left(D^\alpha \theta + R^{\alpha }_{~k}\right)\right. \\ & \left.+\left(D_{\alpha }\sigma_{\beta \gamma }\right)\left(D^{\alpha }\sigma ^{\beta \gamma }\right)\right],
\end{aligned}
\end{equation}
which satisfies \eqref{QNEC} for $\lambda_{\rm Ricci^2} \ge 0$. Here we have used the derivative operator $D_\alpha$ associated with the induced metric $h_{\alpha \beta }$ on the $(d-2)$-dimensional cut of the null congruence, though since $\sigma_{\alpha \beta}$ vanishes at the point of evaluation the Christoffel-symbol terms do not contribute.

It thus remains to discuss the sign of $\lambda_{\rm Ricci^2}$. Since the Ricci-squared term contains a logarithmic divergence for $d=4$ the coefficient of this logarithm should be universal, meaning that it is independent of the renormalization scheme. In fact, as shown in Ref. \cite{Myers:2010jv}, $\lambda_{\rm Ricci^2}$ is related to the ``$c$'' anomaly coefficient by $\lambda_{\rm Ricci^2} = [c/(16\pi ^2)] \ln \Lambda $, and positivity of the stress-tensor two-point function requires $c> 0$. So for $d=4$ all divergent parts of $Q$ satisfy \eqref{QNEC}.

It now remains only to check settings where $\theta $, $\sigma_{\alpha \beta}$, $D_\gamma \sigma_{\alpha \beta}$, and $D_\alpha \theta +R_{k\alpha}$ all vanish so that the QNEC is scheme independent. The story is much the same as for $d=2,3$. From Refs. \cite{Bousso:2015wca,Balakrishnan:2017bjg}, we know \eqref{QNEC} to be satisfied in such cases in flat space \cite{footnote4}, and Ref. \cite{Akers:2017ttv} establishes \eqref{QNEC} under the above conditions in any background for leading-order holographic theories. The similarity of Ref. \cite{Balakrishnan:2017bjg} to the flat-space holographic argument of Ref. \cite{Koeller:2015qmn} again suggests that the techniques of Ref. \cite{Balakrishnan:2017bjg} will allow this arbitrary-background result to be established for general QFTs having nontrivial UV fixed points. And, again, it would be interesting to attempt to extend the free-field light-sheet analysis of Ref. \cite{Bousso:2015wca} in a similar way.

\PRLsection{$d=5$}Three new features arise for $d>4$. The first is that the Gauss-Bonnet term is no longer a topological invariant. As shown in Ref. \cite{Fu:2017lps}, when $\theta $ and $\sigma_{\alpha \beta}$ vanish, it contributes
\begin{equation}
\Delta Q_{\rm GB} = 2\lambda_{\rm GB} \left( 2C_{k\beta \alpha}{}^\beta C_{k\gamma}{}^{\alpha \gamma}
-C_{k\alpha \beta \gamma} C_k{}^{\alpha \beta \gamma}
\right),
\end{equation}
where $C_{abcd}$ is the Weyl tensor of the spacetime metric $g_{ab}$.

The second new feature is that, even for a fixed sign of $\lambda_{\rm Ricci^2}$, the contribution of the Ricci-squared term no longer has a definite sign. Instead, when $\theta $ and $\sigma_{\alpha \beta}$ vanish, Ref. \cite{Akers:2017ttv} gives
\begin{equation}
\label{eq:AnydRicciSquared}
\begin{aligned}
\Delta Q_{\rm Ricci^2} = & 2\lambda _{\rm Ricci^2} \left[ \left(D_\alpha \theta +{R}_{\alpha k}\right)\left(D^\alpha \theta +{R}^{\alpha }_{~k}\right) \right. \\
& \left. +\left(D_{\alpha }\sigma _{\beta \gamma }\right)\left(D^{\alpha }\sigma ^{\beta \gamma }\right)\right. \\
& \left. -\frac{d-4}{2(d-2)}\left(D_\alpha \theta \right)\left(D^\alpha \theta \right)\right].
\end{aligned}
\end{equation}
Finally, the coefficient $\lambda_{\rm Ricci^2}$ becomes power-law divergent and is no longer universal. In particular, it is no longer related to a two-point function and we have found no physical requirement that would fix its sign. In particular, for leading-order holographic theories Ref. \cite{Skenderis:2000in} finds \cite{footnote5}
\begin{equation}
\label{anom}
\lambda_{\rm Ricci^2} = \frac{\ell^2}{(d-2)^2(d-4)},\text{ } \lambda_{\rm GB}=0,
\end{equation}
in terms of the bulk anti-de Sitter scale $\ell$ (which affects the precise definition of the regulator). As pointed out by Eq. (2.32) of Ref. \cite{Cao:2010vj}, when $\theta $ and $\sigma_{\alpha \beta}$ vanish, the Codazzi equation implies a constraint on $D_\alpha \theta $, $D_\gamma \sigma _{\alpha \beta }$, and the spacetime Ricci tensor $R_{ab}$, which is
\begin{equation}
\label{eq:constraint}
\frac{d-3}{d-2}D_\alpha \theta -D_\beta \sigma _{\alpha }^{~\beta }=-R_{\alpha k}.
\end{equation}
If we further impose $D_\gamma \sigma _{\alpha \beta }=0$, Eqs. \eqref{eq:constraint} and \eqref{eq:AnydRicciSquared} yield
\begin{equation}
\Delta Q_{\rm Ricci^2}=-\frac{1}{9}\lambda_{\rm Ricci^2}\left(D_{\alpha }\theta \right)\left(D^{\alpha }\theta \right)\text{, for }d=5.
\end{equation}
Thus, \eqref{QNEC} is violated for any congruence for which $D_\gamma \sigma _{\alpha \beta }$ vanishes and $D_{\alpha }\theta $ does not vanish.

Such congruences indeed exist. To show this, we follow the approach used in Sec. 2 of Ref. \cite{Podolsky:2006du}. This reference shows that a general hypersurface-orthogonal null geodesic congruence can be written as a $v(y^\alpha )=\text{const}$ surface in a spacetime with the metric
\begin{equation}
ds^2=g_{\alpha \beta }\left(dy^{\alpha }+g^{u\alpha }dv\right)\left(dy^{\beta }+g^{u\beta }dv\right)-2dvdu-g^{uu}dv^2,
\end{equation}
where $[\partial /(\partial u)]^a$ is the null normal vector of the congruence and the metric coefficients are arbitrary functions of all the coordinates $(y^\alpha,v,u)$. After defining $q:=\left(\det g_{\alpha \beta }\right)^{1/[2(2-d)]}$ and $\gamma _{\alpha \beta }:=q^2g_{\alpha \beta }$ so that $\det \gamma _{\alpha \beta }=1$, the shear squared and the expansion of the congruence can be written as $\sigma _{\alpha \beta }\sigma ^{\alpha \beta }=\frac{1}{4}\gamma ^{\delta \alpha }\gamma ^{\gamma \beta }\gamma _{\gamma \alpha ,u}\gamma _{\delta \beta ,u}$ and $\theta =-\left(\ln q\right)_{,u}$, respectively. This congruence is shear-free if and only if $\gamma _{\alpha \beta ,u}=0$ everywhere along the congruence. But since this condition does not restrict $q$, it is straightforward to construct an everywhere shear-free congruence for which $\theta$ vanishes at some point $p$ but $D_\alpha \theta|_p \neq 0$. The bare QNEC \eqref{QNEC} is thus violated for such examples.

On the other hand, one should not expect there to be a well-defined procedure for computing $Q$ on length scales smaller than the scale $\Lambda^{-1}$ associated with the cutoff $\Lambda$. So, as in Ref. \cite{Leichenauer:2017bmc}, it is natural to study \eqref{QNEC} only when smeared over scales $L \gg \Lambda^{-1}$. To be precise, we consider an operator $Q_L$ defined by smearing $Q$ over the length scale $L$, fix $L$, and study the asymptotic expansion of $Q_L$ at large $\Lambda$.

From the results above, we need only consider smearing about a point $p$ where $\theta$ and $\sigma_{\alpha \beta}$ both vanish, but if their derivatives fail to vanish then the average Einstein-Hilbert contribution from \eqref{EH} can still be nonzero. For example, expanding $\theta$ in a power series gives
$\theta = 0 + y^\alpha \partial_{\alpha} \theta + \cdots$ so that the average
$\langle \theta^2 \rangle$ of $\theta^2$ is of the order of $ L^2\partial_\alpha \theta \partial^\alpha \theta$ and similarly for $\sigma_{\alpha \beta}$. As a result, for $L^2 \Lambda^2 \gg 1$, the positive-definite Einstein-Hilbert contribution to $Q_L$ will dominate over any possible Ricci-squared or Gauss-Bonnet contribution whenever $\theta$ or $\sigma_{\alpha \beta}$ has a nonvanishing derivative.

It remains only to study cases where these derivatives vanish in addition to $\theta $ and $\sigma_{\alpha \beta}$ themselves. As in Ref. \cite{Leichenauer:2017bmc}, the Codazzi equation then requires
\begin{equation}
0 = R_{k\alpha \beta \gamma} = C_{k \alpha \beta \gamma} - \frac{1}{d-2} \left( R_{k \gamma} h_{\alpha \beta} - R_{k \beta} h_{\alpha \gamma} \right).
\end{equation}
As in Ref. \cite{Akers:2017ttv}, the above results then simplify to yield
\begin{equation}
\begin{aligned}
& \Delta Q_{\rm Ricci^2} + \Delta Q_{\rm GB} \\ = & 2\ell ^2\left(\lambda _{\rm Ricci^2} + 2\frac{(d-3)(d-4)}{(d-2)^2}\lambda_{\rm GB} \right)R_{k\beta} R_k{}^\beta.
\end{aligned}
\end{equation}
Using \eqref{anom}, we then find \eqref{QNEC} to be satisfied by all divergent terms in holographic theories with Fefferman-Graham regulators.

Note that for $d=5$ and with our use of the smeared QNEC the finite terms become relevant only when $\theta $, $\sigma_{\alpha \beta}$, $D_\alpha \theta $, $D_\gamma \sigma_{\alpha \beta}$, and $R_{k\alpha}$ all vanish, in which case the QNEC is again scheme independent. As before, Ref. \cite{Akers:2017ttv} establishes \eqref{QNEC} under the above conditions in any background for leading-order holographic theories and we again expect the method of Ref. \cite{Balakrishnan:2017bjg} to show this in scheme-independent contexts for general QFTs with nontrivial UV fixed points, though it is less clear what physical condition on the regulator could enforce positivity of the divergent curvature-squared contributions to $Q$. It would be interesting to understand whether this is the case for a simple discretization, i.e., for a lattice regulator on a curved spacetime background.

\PRLsection{Summary and discussion}
We argued that for spacetime dimension $d\le 4$ the bare QNEC \eqref{QNEC} should hold for unitary relativistic quantum field theories on general spacetime backgrounds with any choice of null congruence. The essential ingredients were the observations that the anomaly coefficient $c$ is positive for $d=4$ and that the contributions of various local terms to the QNEC in a general background were either computed in Ref. \cite{Fu:2017evt} or can be read off from computations in Ref. \cite{Akers:2017ttv} that were originally performed in order to study the QFC. For $d=3,4$ our only assumption is that the dynamics can depend on the spacetime location only through the background metric (i.e., all other sources are constant scalars) and that the regulator gives any entropy $S$ a positive area-law divergence in the UV.  
Further assumptions required for $d=2,5$ are summarized in Table \ref{tab:1}. In each case, we showed above that all possible divergent terms satisfy \eqref{QNEC}; hence, it follows from Refs. \cite{Fu:2017evt,Akers:2017ttv} that the conjecture holds for leading-order holographic theories. The full conjecture will be established if the proofs in Refs. \cite{Bousso:2015wca,Balakrishnan:2017bjg} can be extended to arbitrary scheme-independent contexts. 

\begin{table}[h]
 \caption{Summary of assumptions (beyond positivity of $S$ and the above restriction on sources) and results by spacetime dimension $d$.}
 \label{tab:1}
 \centering
 \renewcommand{\arraystretch}{1.25}
 \begin{tabular}{|p{0.5cm}|p{3.1cm}|p{2.75cm}|}
  \hline
  {\bf $d$} & {\bf Further assumptions} & {\bf Do divergent terms satisfy \eqref{QNEC}?} \\
  \hline
  $2$ & No operators with free field correlators;  & Yes \\
  & forbidding $\Delta =0$ scalars & \\
  \hline
  $3,4$ & None & Yes \\
  \hline
  $5$ & Holographic QFT with holographic regulator  & Yes when smeared \\
  \hline
 \end{tabular}
\end{table}

Now, the QNEC was introduced in Ref. \cite{Bousso:2015mna} as a consequence of the QFC which concerned theories of quantum gravity in situations where the metric could be described as classical. The QFC imposes the condition
\begin{equation}
\label{QFC}
- S_{\rm gen}'' \ge 0,
\end{equation}
where $S_{\rm gen}$ includes both geometric gravitational entropy and the entropy of all matter fields. The QNEC was argued to follow from \eqref{QFC} by taking the weak gravity limit. Note that \eqref{QNEC} differs from \eqref{QFC} only by the addition of $T_{kk}$ (the expectation value of the $kk$ component of the quantum stress tensor). But the QFC applies only to the total system including both gravity  and matter and, moreover, requires the metric to be on shell. This means that the total partition function is stationary with respect to the variations of the metric. If we now apply the QNEC to the total system, the total $T_{kk}$ is defined by this same variation and thus vanishes; i.e., the QFC is just the special case of our arbitrary-congruence QNEC for which the metric is on shell.  In particular, the hierarchy \eqref{lambdas} used above matches precisely with the conditions under which Refs. \cite{Leichenauer:2017bmc,Akers:2017ttv} argued for a $d=5$ smeared QFC.  Moreover, the analysis of Ref. \cite{Fu:2017lps} suggests that for $d=5$ the QFC will generally be violated when $\lambda_{\rm GB}$ is larger than what would be allowed by \eqref{lambdas}.

It is natural to ask whether our results extend to theories with more general sources. In particular, conformal field theories that admit exactly marginal deformations can be coupled to scalar sources with conformal dimension zero. The presence of such sources allows additional divergent terms beyond those analyzed above. However, for $d=2,3$, the only change is that the leading ($c$-number) term in ${\cal O}_{\rm EH}$ can now depend on such sources. Since the $\dot{\cal O}_{\rm EH}$ term in \eqref{EH} then contributes even at leading order, \eqref{QNEC} must fail for general sources. For $d=3$, the conjecture still holds in the Einstein conformal frame \cite{footnote6}, defined so that ${\cal O}_{\rm EH}$ becomes constant at leading order in $\Lambda$. It would be interesting to find an analogous condition in higher dimensions that again forbids field redefinitions and enforces positivity of the bare QNEC \eqref{QNEC}.

\

It is a pleasure to thank Xi Dong, Tom Faulkner, Tom Hartman, Stefan Leichenauer, Adam Levine, Rob Myers, and Harvey Reall for useful discussions. Z.F. and D.M. were supported in part by the Simons Foundation and by funds from the University of California.


\begin{thebibliography}{99}
	\bibitem{Bousso:2015mna}
	R.~Bousso, Z.~Fisher, S.~Leichenauer and A.~C.~Wall,
	Phys.\ Rev.\ D {\bf 93}, no. 6, 064044 (2016)
	doi:10.1103/PhysRevD.93.064044
	[arXiv:1506.02669 [hep-th]].
	
	\bibitem{footnote1}
	And when enough additional derivatives of $\theta $ and $\sigma_{\alpha \beta}$ also vanish.
	
	\bibitem{Bousso:2015wca}
	R.~Bousso, Z.~Fisher, J.~Koeller, S.~Leichenauer and A.~C.~Wall,
	Phys.\ Rev.\ D {\bf 93}, no. 2, 024017 (2016)
	doi:10.1103/PhysRevD.93.024017
	[arXiv:1509.02542 [hep-th]].
	
	\bibitem{Balakrishnan:2017bjg}
	S.~Balakrishnan, T.~Faulkner, Z.~U.~Khandker and H.~Wang,
	arXiv:1706.09432 [hep-th].
	
	\bibitem{Fu:2017evt}
	Z.~Fu, J.~Koeller and D.~Marolf,
	Class.\ Quant.\ Grav.\  {\bf 34}, no. 22, 225012 (2017)
	doi:10.1088/1361-6382/aa8f2c
	[arXiv:1706.01572 [hep-th]].
	
	\bibitem{Akers:2017ttv}
	C.~Akers, V.~Chandrasekaran, S.~Leichenauer, A.~Levine and A.~Shahbazi Moghaddam,
	arXiv:1706.04183 [hep-th].

    \bibitem{Donnelly:2014fua}
    W.~Donnelly and A.~C.~Wall,
    Phys.\ Rev.\ Lett.\  {\bf 114}, no. 11, 111603 (2015)
    doi:10.1103/PhysRevLett.114.111603
    [arXiv:1412.1895 [hep-th]].

	\bibitem{newfootnote1}	
	Or, at least, an arbitrary state which is sufficiently well behaved in the ultraviolet.
	
	\bibitem{newfootnote2}
	We allow both fundamental scalars (namely, those that appear directly in the action) as well as composite operators such as those that might arise in general operator product expansions.
	
	\bibitem{Wald:1984rg}
	R.~M.~Wald,
	doi:10.7208/chicago/9780226870373.001.0001
	
	\bibitem{footnote2}
	In fact, $\sigma_{\alpha \beta}$ vanishes identically for $d\le 3$, but we carry it along to yield expressions valid in general dimensions. This will be useful below.
		
	\bibitem{Koeller:2015qmn}
	J.~Koeller and S.~Leichenauer,
	Phys.\ Rev.\ D {\bf 94}, no. 2, 024026 (2016)
	doi:10.1103/PhysRevD.94.024026
	[arXiv:1512.06109 [hep-th]].
	
	\bibitem{footnote3}
	Up to a change of sign associated with a reinterpretation of the Ricci-squared term as a ``matter'' term here as opposed to a ``gravitational'' term in Ref. \cite{Akers:2017ttv}; i.e., the $-T_{kk}$ appearing on the right-hand side of their Eq. (3.11) is in fact $+T_{kk}$ in our notation.
	
	\bibitem{Myers:2010jv}
	R.~C.~Myers, M.~F.~Paulos and A.~Sinha,
	JHEP {\bf 1008}, 035 (2010)
	doi:10.1007/JHEP08(2010)035
	[arXiv:1004.2055 [hep-th]].
	
	\bibitem{footnote4}
	Modulo the fact that the exact set of derivatives of $\theta $ and $\sigma_{\alpha \beta}$ that must vanish was not established in Ref. \cite{Balakrishnan:2017bjg}.
	
	\bibitem{Fu:2017lps}
	Z.~Fu, J.~Koeller and D.~Marolf,
	Class.\ Quant.\ Grav.\  {\bf 34}, no. 17, 175006 (2017)
	doi:10.1088/1361-6382/aa80ba
	[arXiv:1705.03161 [hep-th]].
	
	\bibitem{Skenderis:2000in}
	K.~Skenderis,
	Int.\ J.\ Mod.\ Phys.\ A {\bf 16}, 740 (2001)
	doi:10.1142/S0217751X0100386X
	[hep-th/0010138].
	
	\bibitem{footnote5}
	Note that the convention for the Riemann tensor in Ref. \cite{Skenderis:2000in} differs by a sign from the conventions of Ref. \cite{Wald:1984rg}. We use the latter here.
		
	\bibitem{Cao:2010vj}
	L.~M.~Cao,
	JHEP {\bf 1103}, 112 (2011)
	doi:10.1007/JHEP03(2011)112
	[arXiv:1009.4540 [gr-qc]].
	
	\bibitem{Podolsky:2006du}
	J.~Podolsky and M.~Ortaggio,
	Class.\ Quant.\ Grav.\  {\bf 23}, 5785 (2006)
	doi:10.1088/0264-9381/23/20/002
	[gr-qc/0605136].
	
	\bibitem{Leichenauer:2017bmc}
	S.~Leichenauer,
	arXiv:1705.05469 [hep-th].
	
	\bibitem{footnote6}
	The detailed definition of $S''$ turns out to involve the metric volume element $\sqrt{|g|}$. As a result, the QNEC transforms nontrivially under redefinitions of the background fields. In particular, for $d \ge 3$ it transforms nontrivially under conformal rescalings of the metric and cannot hold in a general conformal frame.
\end{thebibliography}
\end{document}